# Generating the right evidence at the right time: Principles of a new class of flexible augmented clinical trial designs

C Dunger-Baldauf[a,1], R Hemmings[b,1], F Bretz[a,c], B Jones[d], A Schiel[e], C Holmes[f]

[a]Statistical Methodology, Novartis Pharma AG, Basel, Switzerland

[b]Consilium Salmonson & Hemmings

[c]Section for Medical Statistics, Center for Medical Statistics, Informatics, and Intelligent Systems, Medical University of Vienna, Vienna, Austria

[d]Novartis UK

[e]Norwegian Medicines Agency

[f]University of Oxford, and The Alan Turing Institute

Correspondence: Chris Holmes (cholmes@stats.ox.ac.uk)

**ABSTRACT**: The past few years have seen an increasing number of initiatives aimed at integrating information generated outside of confirmatory randomised clinical trials (RCTs) into drug development. However, data generated non-concurrently and through observational studies can provide results that are difficult to compare with randomised trial data. Moreover, the scientific questions these data can serve to answer often remain vague. Our starting point is to use clearly defined objectives for evidence generation, which are formulated towards early discussion with health technology assessment (HTA) bodies and are additional to regulatory requirements for authorisation of a new treatment. We propose FACTIVE (Flexible Augmented Clinical Trial for Improved eVidencE generation), a new class of study designs enabling flexible augmentation of confirmatory randomised controlled trials with concurrent and close-to-real-world elements. These enabling designs facilitate estimation of certain treatment effects in the confirmatory part and other, complementary treatment effects in a concurrent real-world part. Each stakeholder should use the evidence that is relevant within their own decision-making framework. High quality data are generated under one single protocol and the use of randomisation ensures rigorous statistical inference and interpretation within and between the different parts of the experiment. Evidence for the decision-making of HTA bodies could be available earlier than is currently the case.

---

[1] These authors contributed equally to this article

**INTRODUCTION**

To support informed decision making by pharmaceutical companies, regulators, health technology assessment (HTA) bodies, payers, patients and physicians, clear descriptions of the benefits and risks of a treatment for a given medical condition should be made available in a timely fashion. The current paradigm is to generate evidence in a sequential manner where at each stage the focus is on one stakeholder and the information they need in order to progress to the next stage. This is inefficient. In this paper we present FACTIVE, a new paradigm where information for regulatory agencies and HTA bodies is generated concurrently via a new class of augmented clinical trial designs. FACTIVE designs study not only patients who are suitable for entry into a conventional Phase 3 randomized controlled trial (RCT) that is conducted under well-controlled conditions but bridges to a broader population, or different experimental conditions, that can be tailored to address a particular HTA question or reflect a particular healthcare system. The proposed framework is different from existing approaches (as we explain later) and allows the generation of the right evidence at the right time, such that key decisions made after Marketing Authorisation (MA) can be made sooner than would otherwise be the case.

**CURRENT STATUS**

Figure 1 (upper panel) summarizes the current evidence generation and decision-making process: Following confirmatory Phase 3 trials, an application for MA of a new treatment is submitted to a regulatory agency. MA is accompanied by a period of discussion and agreement with payers (e.g., HTA bodies, government agencies, medical insurance companies), who will reimburse the cost of the treatment and influence the price at which the treatment should be marketed. The treatment is then placed on the market (❶ in Figure 1). Physicians, health care providers and patients are subsequently informed how the new treatment is positioned in the landscape of already available treatment options (❷) and at some time later (❸) the maximum uptake of its use is achieved. Alongside this, post-authorization trials are conducted to learn more about how the treatment performs in normal clinical practice.

[Insert Figure 1 around here]

The sequential nature of generating the evidence for different stakeholders is immediately visible. The current main driver when designing confirmatory RCTs is to provide sufficient evidence to regulatory agencies of the efficacy and safety of a new treatment in order that it may be granted MA. Additional post-authorization trials provide further evidence of the treatment's effectiveness in a broader patient population under clinical practice conditions. Evidence from such trials, in addition to that provided by the confirmatory trials, is then used to inform further market access and pricing discussions with HTA bodies, taking into account the therapeutic landscape. The post-authorization trials can be RCTs, or open-label extension phases to RCTs, but are commonly undertaken as observational studies. The pre- and post-authorisation experiments are conducted independently of each other, such that if estimates of a particular treatment effect differ between experiments, the reasons for that difference cannot be determined with certainty.

Multiple initiatives have attempted to streamline the evidence generation that will support regulator, HT assessor, payer, prescriber, and patient decision-making [Eichler et al. 2016; Califf et al. 2016; Ray et al. 2022]. Efforts in this direction must address the fact that different stakeholders have different questions to address, including the benefit-risk of an intervention within a specific target population vs the cost-effectiveness of an intervention including societal perspectives and a specific healthcare budget. Decision-making by the European Medicines Agency (EMA) is centralised on behalf of the European Union (EU) whereas decisions made by one or more country specific HTA bodies are national or local. Different stakeholders will also identify different sources of uncertainty and evidence gaps they want to see addressed, preferably during the clinical evidence generation phase, or at least post-authorisation. This paper focuses on evidence generation to meet the needs of regulatory agencies and HTA bodies, as they make the two initial and most critical public-sector decisions to determine patient access to medicinal products. The EU is taken as a jurisdiction for illustration, though the benefits of the approach described can be realised more generally.

A point of particular focus for discussions on streamlining evidence generation has been the design and conduct of RCTs. All stakeholders recognise the high internal validity of this experimental design: the fact that reliable treatment effect estimates for well-defined research questions can be provided through a design where experimental conditions are controlled and well-understood. Indeed, deriving a reliable estimate and being able to interpret the magnitude of treatment effects in the context of experimental conditions that are documented and understood are the attributes of the RCT that make it the "gold-standard". In addition, all stakeholders understand that there can be a scientific basis to generalise, or extrapolate, inferences from a clinical trial dataset to a broader patient population or clinical context, though the basis for extrapolation (e.g., other clinical trial data, pharmacological understanding of the mechanism of action, pharmacological modelling), the extent of the extrapolation (to what proportion of the target population does the extrapolation apply) and the type and strength of evidence needed to support extrapolation is not documented and hence not unified for benefit-risk vs cost-effectiveness decisions. Importantly, an RCT design in which the experimental conditions are controlled too tightly can leave all stakeholders questioning its external validity, i.e., the applicability of the trial results to the intended patient population and therapeutic use in clinical practice.

External validity of a trial is assessed in relation to its inclusion and exclusion criteria (vs the population indicated for clinical practice) and its experimental conditions (vs the therapeutic use expected in clinical practice) such as the outcome variable or comparator, permitted concomitant medications or combinations of treatments. The different mandates for regulators and HT assessors can also lead to different clinical outcomes being prioritised for the assessment of efficacy or effectiveness with a consequence for the periods of treatment and follow-up that are of interest, and potentially different treatment effects of interest (i.e., estimands) [ICH, 2019; Remiro-Azócar, 2022]. Importantly, whereas a given clinical development programme will be targeted towards a centralised regulatory approval, each HT assessment of the applicability of the trial results to their specific national or local jurisdiction might differ, for example, in relation to other products that are/are not reimbursed and used locally, or the precise target population for which cost-effectiveness can be justified. Concerns over the external validity of a particular RCT does not represent a fundamental flaw in that study design and conduct, only that the specific trial design cannot directly address the needs of a specific HTA body. Section 1 of the Supplementary Material gives additional discussion on external validity and extrapolation.

The result of these dynamics is that a regulator might authorise a product, perhaps with post-authorisation evidence generation to address identified uncertainties, whilst an HT assessor might not feel fully informed about how the product will impact their specific healthcare system and budget, and whether a positive decision on cost-effectiveness can be justified. To address a broader set of stakeholder needs, RCTs might be made larger and/or longer and/or less well controlled in respect of patient population and experimental conditions. In addition, different endpoints or multiple comparators might be used. In reality, however, complementary sources of evidence generation are more efficiently used to provide answers to general and specific questions from HTA bodies. An often-overlooked fact is that the information required to strengthen the external validity can be generated concurrently with the trial data. Additional evidence is not necessarily generated to replicate trial results, rather additional data can explore the effects of treatment beyond the patient population and experimental conditions of the RCT. This paper discusses an experimental design that provides these additional data and seeks to deliver information to all stakeholders in a timely manner, preserving efficient evidence generation for each stakeholder and promoting a methodologically robust approach in an experimental design where different parts are no longer independent.

## FLEXIBLE AUGMENTION TO GATHER THE RIGHT EVIDENCE AT THE RIGHT TIME

We argue that the understanding of the relative effectiveness and time to peak uptake of a new treatment can be enhanced, without compromising safety, by generating additional rigorous evidence throughout the confirmatory development process. To do so we propose FACTIVE, a new class of augmented RCT designs aimed at widening the evidence base of traditional RCTs. The lower panel of Figure 1 summarizes the potential impacts of using such an augmented RCT design (which we describe below): Discussions with HTA bodies are better informed and shortened, along with a potentially greater maximum uptake of treatment use.

A key feature of the new paradigm is that augmentation is wrapped around a conventional RCT that is designed in the usual manner to focus on treatment efficacy and safety in a controlled experimental environment. The consequence is that the core RCT, or RCTs, which form the pivotal evidence for a regulatory approval, is ring-fenced. A cross-disciplinary team can consider the specific objectives, subsequent design criteria and the timing for augmentation with real-world elements, including consideration of market value and patient heterogeneity as well as early evidence of efficacy and safety in the core RCT. The augmentation can resolve uncertainties that could not be achieved by simple improvements to the RCT.  For example, providing some insights into multiple, different combinations of active comparators are classic examples of HTA requirements that might dramatically increase the size, duration and cost of a confirmatory RCT.

We set no limitations to the scope of research questions that can be addressed through augmentation. The questions to be answered by augmentation may be general: to provide estimates of treatment effects in the target population reflecting routine clinical care and under conditions reflecting clinical practice; to facilitate data integration with an existing external data source, by augmenting the RCT with subject eligibility criteria and conditions matched to the external resource; or, alternatively, targeted to obtain complementary information on a specific relaxation of an inclusion/exclusion criteria or different methods of outcome assessment. As an example of a specific question, consider a sponsor needing to address differences in national treatment guidelines regarding a background therapy. Instead of including patients on various background therapies, the core RCT could be conducted on one background therapy. Information as add-on to various background therapies (including the one in the core RCT) could be generated in the augmented part, whilst the patient population and experimental conditions remain otherwise similar, to establish that there is no impact of the background therapy or to characterise the impact that changing background therapy might have on the treatment effect that was observed in the RCT.

FACTIVE supplements the evidence provided by the core RCT for MA through an increased sample size with additional information from close-to-real-world (cRW) elements carefully selected according to safety, feasibility and, critically, the outstanding questions to be answered. There are established mechanisms for sponsors to interact with regulators (e.g., Scientific Advice procedures in the EU) to understand preferences and standards for a future application. Early dialogue with HTA bodies can provide valuable input about the context in which a treatment might be assessed, and important considerations that are not addressed in the core RCT. For example, PICO (Population, Intervention, Comparator(s), Outcomes) provides a framework to compare the evidence being generated to the question of interest for the HTA and, while the PICO can change over time, it can be used to explore whether limitations to the core RCT evidence should be anticipated due to, e.g., missing sub-populations, differences in treatment algorithms, or preferences for other comparators. Identifying potential evidence gaps can then inform the purpose, and consequently the design, of the augmentation. Again, the augmentation is not designed to serve as confirmation of the RCT evidence but represents a basis to provide data, or to bridge the RCT data, to the evidence requirements of other stakeholders.

**THE FACTIVE DESIGN**

The FACTIVE design is to generate evidence rigorously and contemporaneously with high-quality information obtained through randomisation. A sketch of the augmented design is given in Figure 2.

Two types of patients are identified in FACTIVE: Those eligible for the core RCT (green) and those who are from a broader population (blue). In addition, two experimental settings are identified, the RCT conditions and the alternate cRW experimental settings (e.g., those closer to clinical practice). Both types of patients are first randomized to be studied under RCT treatment conditions or under cRW treatment conditions. The RCT-eligible patients who are studied under RCT conditions form Part A of the design, the core RCT used for regulatory submission. Part B is comprised of additional RCT-eligible patients (green) and those from the broader population (blue) randomized to cRW treatment conditions and patients from the broader population randomized to RCT treatment conditions. Within each part of the design, patients are randomized to either the experimental treatment or control. This allows all conventional RCT analyses for authorization purposes to be conducted with the evidence generated in Part A, supplemented by other analyses addressing specific questions in Part B.

[Insert Figure 2 around here]

The nested structure facilitates rigorous statistical analyses for causal effects of interest, as explained in Section 2 of the Supplementary Material. The augmented design makes it possible to estimate and compare treatment effects, and effect changes, across the four combinations of subject eligibility (RCT eligible patients versus broader population) and treatment conditions (RCT versus cRW treatment conditions).

The augmented design is complementary to existing trial designs which look to combine RCT and cRW elements such as seamless Phase 3/4 designs [Eichler, 2010] and clinical trials using external control information [Schmidli et al., 2020]. The distinguishing feature of FACTIVE is the collection of data on randomized cRW elements under the new treatment, before regulatory approval, and concurrently with corresponding RCT data, thus enhancing the available evidence. Note that the left-hand side of Part B in Figure 2 could be implemented as a pragmatic trial. It fulfils the pragmatic study criteria for randomized studies [Zuidgeest, et al, 2017] under clinical practice treatment conditions, of patients expected in routine clinical care, and of assessments meaningful to patients/physicians. However, FACTIVE is broader than conventional pragmatic trial designs: while concurrency reduces the sources of time-related bias, the core RCT in Part A together with this left-hand pragmatic part does not allow a rigorous comparison of cRW and RCT treatment effects beyond the RCT-eligible patients. The starting point for FACTIVE follows the design of the core RCT, which is subsequently assessed for cRW augmentation. The additional part on the right-hand side, where patients from the broader population are treated under RCT treatment conditions, then provides a more comprehensive understanding of treatment effects.

There can be flexibility also in the design of Part B. Note that removing the cRW components from the augmented design simply returns the original RCT. Note also that the design elements are flexible: If few inclusion/exclusion criteria are used in the RCT, the concept of a broader population might be redundant. Likewise, if the experimental conditions used in the RCT are close to clinical practice it might not be necessary to examine the treatment under alternative conditions. It is not even necessary for the same control arm to be used in both parts of the experiment, though then some of the benefits of the design are lost. If different controls are used in Part A and Part B, network meta-analysis [Dias et al., 2018] is one method that can be used to bridge from one control to the other using relevant external data. To justify the proportion of patients from the RCT-eligible and the broader patient population along with the respective sample sizes, whilst one approach might be to adequately power a comparison of treatment vs control in cRW conditions, other approaches are conceivable. Returning to the example above of investigating the effects of an experimental treatment on different background treatments, the amount of information to be generated might be thought of as similar to generating evidence across subgroups to enable an assessment of consistency (CHMP, 2019). As stated above, the objective is not to replicate findings from the core RCT. In particular when creating evidence to bridge from the RCT, or to give confidence in the external validity of the RCT results, precision of estimates might be more important and a better basis for planning than tests of statistical significance.

## DISCUSSION

The development of new treatments is a continuous learning process where evidence collected in early phases contributes to decisions made in later phases and where techniques commonly used in exploratory development can continue to be used on confirmatory data (e.g., clinical pharmacology modelling) [Sheiner, 1997]. Even in confirmatory Phase 3 trials, modifications can be made as knowledge grows during their execution as, for example, in an adaptive trial with dose selection [CHMP, 2007; FDA, 2019], or relaxation of an inclusion criterion after preliminary safety data have been reviewed. Augmenting confirmatory RCTs (Part A in Figure 2) with tailored cRW elements (Part B in Figure 2) can be considered a natural extension of this process. Whilst FACTIVE will not be a suitable approach to evidence generation in every programme, we think all programmes can benefit from a discussion on the merits of structured and concurrent evidence generation beyond the core RCT. The key is not to implement a fixed design or to follow a checklist, but to carefully consider uncertainties or evidence gaps that are critical to address through evidence generation beyond a well-designed confirmatory RCT.

In contrast to the current practice, FACTIVE offers more timely evidence generation and an increased potential to investigate and quantify different modifiers of the treatment effect. Under the current approach, an estimated treatment effect might differ between pre- and post-authorization experiments for reasons that are often not fully understood. This might be due to changes in care over time, changes in patients recruited, perhaps due to lack of or different choice of control arm or changes in experimental conditions, including methods or timings of assessments applied, adherence to treatment or use of concomitant treatments, the choice of investigational sites, etc. Alternatively, estimands [ICH, 2019] may differ (intentionally or unintentionally), such that the experiments address different clinical questions of interest. Compared to uncontrolled observational studies, however, the collection of randomized data based on cRW elements concurrently with the core RCT data allows for a unified statistical analysis with nested models, without the biases inherent when comparing or integrating data from different experiments. If cRW data are collected from observational studies after the RCT then assumptions on the impact of potential confounders (untestable with the data to hand) are needed for a joint statistical analysis to proceed.

The time to initiate Part B would depend on having clarity on the research questions to be addressed and perhaps on accumulating evidence from Part A, for example into the safety of exposing a broader patient population or relaxing the experimental conditions. However, an obvious disadvantage of staggering the start of Part B would be the reduction or elimination of the overlap in time between the two parts. Only with the two parts of the experiment conducted concurrently, enabling randomisation of patients between the different parts of the experiment and between experimental conditions, can the full strength of the design and insights into effect modification be leveraged.

Of course, there may be barriers to the adoption of FACTIVE. We acknowledge our ideas have the potential to be disruptive and may not be taken up easily by all stakeholders. However, in order to break the current linear thinking, changes in mindset are needed, which will take concerted efforts by all stakeholders. The most important change is the move towards an active consideration of the right time for evidence generation, taking account all available information. The optimal timeline will not always be the conventional sequential one (Figure upper panel). A paradigm shift is needed towards understanding that complementary evidence can, and where possible should, be generated in parallel but without the intention to change the framework for regulatory or HTA assessment. FACTIVE aligns with initiatives such as the EMA/HTA parallel Scientific Advice procedure that have promoted early consideration of the different stakeholder needs and post-authorisation evidence generation. Maximising the benefits of this requires full engagement of drug developers bringing all relevant disciplines into discussions from an early stage.

Having the RCT and cRW data available simultaneously to both regulatory agencies and HTA bodies is a new paradigm. How much weight each party should give to the two types of evidence will be context dependent and specific to the designs and results of the different experiments in the context of the totality of evidence generated. It is important that stakeholders are well versed in critically appraising the strengths and weaknesses of different experimental approaches. Optimal stakeholder decision-making cannot be based on the rhetoric that RCTs have inadequate external validity and real-world experiments have inadequate internal validity. Each experiment should be judged on its own merits with an understanding that it is possible to generalise results from an RCT if potential effect modification is understood and that it is possible to interpret evidence of benefit from "real-world" experiments when well-conducted and reported. Randomisation is a strength in the cRW experiment even if the experimental conditions are less well controlled.

As with any design, there is the potential for misuse. The potential to generate data in a broader patient population under cRW conditions should not be used as a reason for the sponsor to tighten the conditions and lessen the external validity of a core RCT. On the other hand, the potential to generate additional evidence in a timely manner should not mean that the demands of regulators and HTAs increase. We reflect further on the EMA/HTA parallel Scientific Advice procedure, whereby despite a tri-partite conversation that ranges wider than only a regulator's or a HTA's individual needs, each stakeholder gives advice to drug developers in a way that reflects and is confined to their own respective legislative basis and mandate. The parallel to FACTIVE is the design of fit-for-purpose RCTs with a vehicle to generate evidence on additional research questions so that stakeholders can make timely decisions, but without altering the established decision-making frameworks. Having clear objectives for the concurrent evidence generation will help to clarify the relevance of the evidence to the different stakeholders.

In conclusion, through FACTIVE we propose a class of augmented randomized controlled trial designs involving the concurrent collection of RCT and cRW data based on close-to-real-world elements in a way that is optimally and sequentially organized by generating the right evidence at the right time. This approach retains the distinctiveness of the core RCT and the questions it seeks to answer from additional research questions aimed at investigating the performance of the treatment in a broader population under cRW conditions or to investigate one or more critical, specific limitations to the RCT design.

Through the use of randomization, FACTIVE ensures that the data on cRW elements are of similar high quality to conventional RCT data and available in closer proximity to the time of MA and able to inform initial HTA discussions, accelerating the journey of the novel treatment from discovery to patient. Discussions with regulators and HTAs can take place either simultaneously or in rapid sequence after completion of the core RCT with the aim that treatments can be made available to patients earlier. In addition, information available to physicians, health care providers and patients will be enhanced by the evidence provided by the augmented design. This could lead to a greater awareness of the benefits of the treatment, leading to its greater use in the community. Post-authorization trials would still be required to collect data on post-authorization aspects of the use of the new treatment, but these are likely to be fewer and smaller in number given the high quality cRW evidence already available.

**Figure 1**

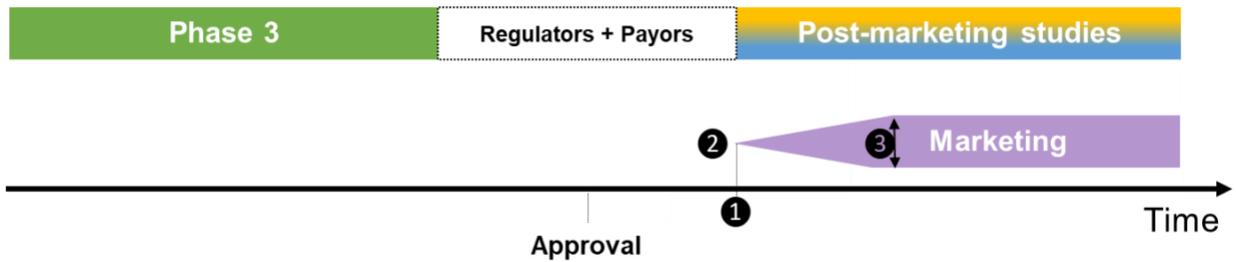

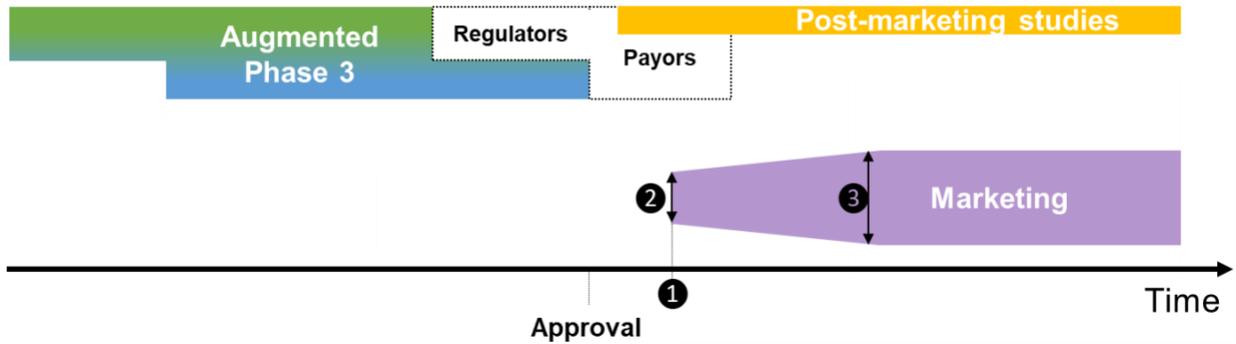

Figure 1. Current (upper panel) and proposed (lower panel) timing of cRW data element collection in Phase 3 treatment development. ❶ First availability of new treatment to patients; ❷ Increasing uptake of new treatment use; ❸ Maximum uptake of new treatment use

**Figure 2**

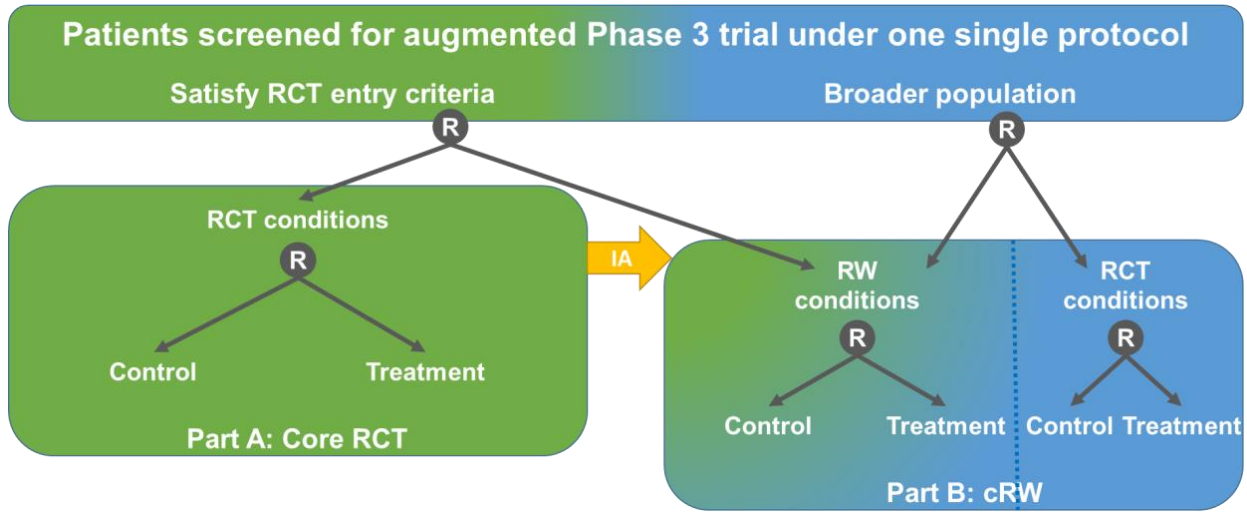

Figure 2. FACTIVE. Patients screened for the augmented Phase 3 trial are recruited from the target population expected to be treated post-authorization. These are either RCT-eligible (green) or from a broader patient population (blue), excluding those patients with, for example, a safety risk. RCT-eligible patients are randomized either to Part A (the core RCT) or to Part B, where they are treated under cRW conditions. In both parts they are additionally randomized to experimental treatment or control. Patients from the broader population are assigned to Part B only and are randomized to be treated under cRW conditions or under RCT conditions. In both situations they are then randomized to experimental treatment or control. In Part B, the proportions of green and blue patients treated under cRW conditions need to be agreed beforehand (e.g., to match epidemiology). RCT/cRW conditions define RCT/cRW design elements such as visit schedule, administration of treatment, monitoring, but exclude specifics about the patient population. The time to initiate Part B could be made dependent on accumulating evidence from Part A at an interim analysis (IA) as illustrated by the vertical offset and the arrow in yellow between Part A and B.

# 1 External validity and Extrapolation

Clinical trial design aims to strike a balance between providing data that with internal validity, to reach a robust causal inference conclusion on treatment effects and with external validity so that results can be deemed applicable to a broad target population and routine clinical practice. However, these priorities can compete, such that attempts to increase internal validity come at the cost of reducing external validity. Where trials have questionable external validity in the context of a particular regulatory or HTA assessment, the need to evaluate evidence in a contextualized manner explicitly requires an extrapolation to under-represented patient subsets or the different conditions of routine clinical practice. The issues and considerations involved are complex and nuanced. As we assert in the main paper: optimal stakeholder decision-making cannot be based on the rhetoric that RCTs have inadequate external validity. We illustrate the nuances below, presenting examples of issues related to the design and conduct of RCTs and exploring whether each issue truly presents a problem in respect of external validity for sponsor and stakeholder decision-making. Confirmatory trials are generally run under well-controlled clinical conditions so that treatment effect estimates can be well understood and interpreted. Inclusion / Exclusion criteria give a clear identification and description for a study population and the protocol outlines the conditions of the experiment under which the treatment conditions are investigated and compared. To assist with interpretation, full information is available on, e.g., adherence, assessment methods and schedules, use of concomitant treatments and other elements that might moderate the estimated treatment effect. These are strengths of the RCT. As the heterogeneity of either a trial population or the trial conditions increases, the overall estimates of treatment effects become harder to interpret. Where the trial population or trial conditions include multiple factors that modify the treatment effect, an overall estimate of effect might not neither be interpretable nor meaningful. Still, a trial that is more homogenous in either regard is not necessarily a problem in respect of external validity. For example, if a drug shows a meaningful effect as an add-on to Drug A, there might exist a sound scientific rationale to consider that a meaningful effect would also exist when the drug is used as an add-on to Drug B. Then external validity is not compromised even if the add-on to Drug B is not tested in the trial. Similarly, patients with impaired renal function or certain concomitant medications might be excluded from an RCT on the basis that insufficient information is presently available to ensure patient safety. Evidence from clinical pharmacology studies might then be available to confirm similar PK/PD responses or absence of drug-drug interactions such that the results from the experiment can also be applied to these specific patient subgroups despite their exclusion from the trial. Again, external validity is not compromised. More generally, RCTs tend to include patient populations selected according to detailed inclusion and exclusion criteria. However, these criteria can be specified for various reasons, with greater or lesser impact on the generalisabilty of the study results – in terms of whether beneficial treatment effects exist and their magnitude, and whether risks of treatment might not have been identified or adequately characterised. Some examples are given in Table 1.

Table 1: Reasons for RCT entry restriction and the impact on generalisability

| **Reason for restriction** | **Example** | **Impact** |
|---|---|---|
| Known safety issues of the experimental product. | Based on information derived from earlier phase trials or trials of products of similar pharmacology. | Yes. Additional evidence generation would be required if it is ever intended to extend use into the population to be excluded. |
| Uncertainties in relation to the safety of the experimental compound. | Ongoing clinical pharmacology studies in patients with renal impairment, with consequences for exposure not yet known. | Yes Will need to justify inclusion of these patients in the target population through additional evidence generation. |
| Known safety issues of reference product. | A contraindication of the control arm will influence trial's inclusion/exclusion criteria | Perhaps none, if it can be argued that if those aspects will not impact the efficacy or safety of the experimental treatment. |
| Enrichment. | More severe / rapidly progressing disease in order to accumulate events / deterioration more rapidly. | Perhaps none, if there is justification that relative effects will be similar in a broader patient population and that relevant benefits are preserved in absolute terms. |
| Increasing assay sensitivity. | Restricting the trial to patients that are stable in terms of symptoms and / or medications in attempts to more efficiently isolate the effects of the experimental treatment. | Justifications would need to be available that relevant (even similar) effects sizes would be observed when conditions were relaxed, otherwise additional evidence generation could be required. |

## 2 Estimands and estimation in the augmented design

### 2.1 Introduction

In the following we consider an augmented design as illustrated in Figure 2, consisting of two parts run under a single protocol. Patients screened for the augmented Phase 3 trial are recruited from the target population expected to be treated post-authorisation. These are either RCT-eligible or from a broader patient population, excluding those patients with, for example, a safety risk. RCT-eligible patients are randomized to Part A (the core RCT) or to Part B, where they are treated under cRW conditions. In both parts they are then randomized to control or drug. Patients from the broader population are assigned to Part B only and are randomized to be treated under cRW conditions or under RCT conditions. They are then randomized to control or drug. The time to initiate part B could be made dependent on accumulating evidence from part A, for example using Bayesian decision rules.

### 2.2 Notation

- $Y$ denotes the treatment response, i.e., the outcome of interest.
- $T \in \{0, 1\}$ is a binary treatment indicator, $T = 1$ for the experimental treatment (i.e. drug), $T = 0$ for standard-of-care or placebo (control).
- $X \in \mathcal{X}$ are patient-specific covariates
- $C$ defines a subset of $\mathcal{X}$ that relates to the RCT eligibility criteria, $C'$ is the complement set defining RCT ineligibility (i.e., the broader population), with $C \cap C' = \emptyset$, $C \cup C' = \mathcal{X}$, such that $\mathbb{1}(x_i \in C)$ indicates that patient $x_i$ is eligible for the RCT.
- $P \in \{0, 1\}$ indicates the conditions under which the treatment is administered, $P = 1$ for the strictly controlled RCT conditions, and $P = 0$ for the more flexible cRW conditions.
- $\Delta(A)$ indicates a conditional average treatment effect, given conditions $A$,

$$\Delta(A) \equiv E[Y \mid T = 1, A] - E[Y \mid T = 0, A].$$

We suppose that interest is in the average treatment effect (ATE), $E[Y \mid T]$, and conditional average treatment effects (CATEs), $E[Y \mid T, A]$ for conditions $A$. Treatment effects in the treated (ATT) might also be of interest, e.g., if a control group under cRW conditions is not available.

### 2.3 Estimands

The estimand framework (ICH, 2019) defines five attributes of an estimand determined by the scientific question which the trial aims to answer. These attributes are treatment, population, variable of interest, handling of intercurrent events, and the summary measure. The treatment attribute might not only be determined by the dose and regimen of the new drug, but also by protocol conditions with strict schedules and monitoring in the RCT vs more flexible treatment conditions in the concurrent cRW part.

To lay out the concept we assume that the five attributes are already defined corresponding to the specific question and that the strategy for handling intercurrent events is determined.

Intercurrent events (ICE) observed in the RCT (Part A) could differ from those in the cRW (Part B). For example, 'an assessment is not performed as it is not needed for the physician's treatment decision' could be an ICE in Part B but would not occur in Part A.

The use of an augmented design (Figure 2) supports unbiased estimation of the following estimands:

1. Treatment effect under the RCT treatment conditions for RCT-eligible patients (i.e., in Part A)

$$\begin{aligned} \theta_1 &= \Delta(X \in C, P = 1) \\ &= E[Y \mid T = 1, X \in C, P = 1] - E[Y \mid T = 0, X \in C, P = 1]. \end{aligned}$$

This is the usual estimand (and the only available estimand) from a conventional RCT. This confounds $X \in C$ and $P = 1$, meaning that possible treatment interactions with population characteristics, $X$, and the treatment conditions, $P$, cannot be differentiated further. Using an augmented design we are also able to estimate

$$\begin{aligned} \widetilde{\theta_{\tilde{1}}} &= \Delta(P = 1) \\ &= E[Y \mid T = 1, P = 1] - E[Y \mid T = 0, P = 1], \end{aligned}$$

the CATE conditioned on the RCT treatment conditions (independently of $X$). As the entire study population is comprised of RCT eligible ($X \in C$) and RCT ineligible patients ($X \in C'$), combined treatment effects are linear combinations of the CATE conditioned on $X \in C$ and the CATE conditioned on $X \in C'$. The respective weights sum to 1 and need to be set according to the scientific question. Study populations could be weighted equally, for example, or according to their size, or to a target population.

The design (Figure 2) is flexible and can be tailored to answer the scientific questions of interest. This could mean that certain design elements would not be implemented. Writing the estimands in terms of CATEs conditioned on $X \in C$ or $(X \in C')$ clarifies what can still be estimated with the available design elements. For example, the purpose of treating RCT-ineligible patients under RCT treatment conditions might not seem obvious. The estimand based on this design element is given by $\theta_7$ below. Laying out the estimands as linear combinations of the CATEs conditioned on the study population(RCT-eligible or not) clarifies where e.g. this design element contributes.

Here we have

$$\widetilde{\theta_{\tilde{1}}} = w_{1\tilde{1}}\theta_1 + w_{2\tilde{1}}\theta_7,$$

where $w_{ij}$ are weights for weight number $i$ and estimand number $j$ ($i$=1,2; $j$ =1,$\tilde{1}$,...,8). Estimand $\theta_7$ is defined below.

2. Treatment effect in the treated between treatment conditions

$$\theta_2 = E[Y \mid T = 1, P = 1] - E[Y \mid T = 1, P = 0].$$

Note that $\theta_2$ isolates the effect of the treatment conditions on the treated and is independent of $X$. We can write estimand $\theta_2$ as a linear combination of the effects in the RCT-eligible and RCT-ineligible patient populations

$$\begin{aligned} \theta_2 &= w_{12} \left( E[Y \mid T = 1, X \in C, P = 1] - E[Y \mid T = 1, X \in C, P = 0] \right) \\ &+ w_{22}(E[Y \mid T = 1, X \in C', P = 1] - E[Y \mid T = 1, X \in C', P = 0]). \end{aligned}$$

3. Difference of treatment effects between treatment conditions

$$\begin{aligned}
\theta_3 &= \Delta(P=1) - \Delta(P=0) \\
&= (E[Y \mid T=1, P=1] - E[Y \mid T=0, P=1]) - (E[Y \mid T=1, P=0] - E[Y \mid T=0, P=0]) \\
&= E[Y \mid T=1, P=1] + E[Y \mid T=0, P=0] - E[Y \mid T=0, P=1] - E[Y \mid T=1, P=0].
\end{aligned}$$

Note this is independent of $X$.

As laid out above, the estimand $\theta_3$ is the difference of the treatment effect under RCT treatment conditions and the treatment effect under cRW treatment conditions. The treatment effect under RCT conditions is a linear combination of $\theta_1$ ($X \in C$) and $\theta_7$ ($X \in C'$). The treatment effect under cRW treatment conditions, $\theta_8$ (see below), can be displayed as a linear combination of the study population CATEs.

4. Heterogeneous treatment effect in the treated under RCT treatment conditions

$$\theta_4 = E[Y \mid T=1, X \in C, P=1] - E[Y \mid T=1, X \in C', P=1].$$

5. Effect of treatment conditions on the treated for RCT eligible patients

$$\theta_5 = E[Y \mid T=1, X \in C, P=1] - E[Y \mid T=1, X \in C, P=0].$$

6. Treatment conditions effect on the treated for RCT ineligible

$$\theta_6 = E[Y \mid T=1, X \in C', P=1] - E[Y \mid T=1, X \in C', P=0].$$

7. Treatment effect under RCT treatment conditions for RCT ineligible patients

$$\begin{aligned}
\theta_7 &= \Delta(X \in C', P=1) \\
&= E[Y \mid T=1, X \in C', P=1] - E[Y \mid T=0, X \in C', P=1].
\end{aligned}$$

8. Treatment effect under the cRW treatment conditions

$$\begin{aligned}
\theta_8 &= \Delta(P=0) \\
&= E[Y \mid T=1, P=0] - E[Y \mid T=0, P=0].
\end{aligned}$$

Note this is independent of $X$.

In terms of CATEs conditioned on study population,

$$\begin{aligned}
\theta_8 &= w_{18}(E[Y \mid T=1, X \in C, P=0] - E[Y \mid T=0, X \in C, P=0]) \\
&+ w_{28}\left(E[Y \mid T=1, X \in C', P=0] - E[Y \mid T=0, X \in C', P=0]\right).
\end{aligned}$$

We can now answer the question posed at the beginning of this subsection. The treatment effect under RCT treatment conditions for RCT ineligible patients $\theta_7$, or the first summand for the treated, contributes to estimands $\widetilde{\theta_{\tilde{1}}}$, $\theta_2$ (ATT), $\theta_3$, $\theta_4$ (ATT), and $\theta_6$ (ATT). In addition, estimand $\theta_7$ provides a direct answer to potential questions of practising physicians about the therapeutic benefit of RCT-ineligible patients under RCT treatment conditions and could support recommendations in clinical practice. If RCT-ineligible patients are not treated under RCT conditions by design, comparisons of treatment conditions will be restricted to RCT-eligible patients.

## 2.4 Estimation

To begin we shall assume that the full augmented design is applied, that the estimand framework is in place, and that the relationship between the outcome Y and P, T, RCT eligibility and possibly patient-specific covariates $X \in \mathcal{X}$ can be modelled by linear regression.

In order to estimate $(\theta_1, \ldots, \theta_8)$ we introduce a set of binary indicators, $z_i = (z_{i1}, z_{i2}, z_{i3})$, for each of $i = 1 : n$ patients in the augmented trial.

- $z_{i1} = P_i$, where $P_i$ is the treatment conditions assigned to the $i$'th patient.
- $z_{i2} = \mathbb{1}(x_i \in C)$ indicates patient $i$ was eligible for the RCT.
- $z_{i3} = T_i$ indicate the treatment received by the $i$'th patient.

Assuming $n_1$ patients are enrolled to be studied under the RCT conditions ($P = 1$), and $n_0$ patients are enrolled to be studied under the cRW conditions ($P = 0$), with $n = n_1 + n_0$, then we can form an $(n \times 3)$ indicator matrix, $Z$, with row vectors $z_i$ relating to the $i$'th patient. The $Z$ matrix can be extended to include covariates $X \in \mathcal{X}$. In this case, all of the estimates, $(\widehat{\theta}_1, \widehat{\theta}_{\tilde{1}}, \ldots, \widehat{\theta}_8)$, can be obtained from applying standard ANOVA or ANCOVA techniques, according to Section 2.3, with associated confidence intervals, tests of hypotheses, and p-values.